\documentclass[11pt,a4paper]{article}

\pdfoutput=1

\usepackage[margin=1in]{geometry}
\usepackage{times}
\usepackage{graphicx}
\usepackage{booktabs}
\usepackage{hyperref}
\usepackage{natbib}
\usepackage{amsmath}
\usepackage{xcolor}
\usepackage{multirow}
\usepackage{tabularx}
\usepackage{float}
\usepackage{amssymb}
\usepackage{rotating}
\usepackage{longtable}
\usepackage{pdflscape}

\hypersetup{
  colorlinks=true,
  linkcolor=blue,
  citecolor=blue,
  urlcolor=blue
}

\title{\textbf{EvalLoop: A Methodology for Evaluation-Driven Iterative Improvement of Business AI Systems}}

\author{
  Kenneth Benavides \quad Josh Fleischer \quad Danti Chen \\
  \textit{Robert Half} \\
  {\tiny\texttt{(kenneth.benavides, josh.fleischer, danti.chen)@roberthalf.com}}
}

\date{}

\begin{document}
\maketitle

\begin{abstract}
Teams deploying large language models in business contexts need evaluation systems, yet most treat evaluation as static model selection: run benchmarks, rank models, deploy the winner. This framing misses evaluation's primary value for production systems---diagnosing \textit{why} a system underperforms and guiding \textit{what to fix}.

We present \textbf{EvalLoop}, a methodology for evaluation-driven iterative improvement. EvalLoop organizes evaluation around three mechanisms: (1)~dimensional metric grouping that decomposes quality into business-relevant dimensions enabling orthogonal failure diagnosis; (2)~failure mode classification that categorizes \textit{why} outputs fail within weak dimensions, bridging diagnosis to action; and (3)~a structured iteration workflow where each evaluation run varies one system variable and compares dimensional profiles before and after.

We validate EvalLoop through a case study on sales intelligence briefing generation (10 models, 3 providers, 18 metrics, 5 dimensions, 3 iterations). Dimensional diagnosis identified that 69\% of hallucination failures were prompt-induced interpretation errors---invisible in aggregate scoring. A targeted prompt fix improved the best model from 82.6\% to 94.6\% overall, with improvement concentrated in diagnosed dimensions (Content Accuracy +16.8pp, Synthesis Power +26.4pp). An undirected configuration change in a prior iteration produced zero impact, illustrating the cost of iterating without diagnosis. We additionally demonstrate that dimensional profiling enables deployment-specific model selection, and that a one-time blind human gate on a finalist panel (4 models, 16 cases) confirms dimensional rankings while resolving multi-criteria deployment trade-offs---a 94\% reduction in human review burden compared to evaluating the full design.

EvalLoop is packaged as reusable artifacts (playbook, agent specification, template repository) for adoption by other teams.
\end{abstract}

\section{Introduction}

Organizations deploying large language models (LLMs) in business contexts invest heavily in evaluation---comparing models on quality metrics, measuring compliance with domain requirements, and validating outputs against business rules. Yet the dominant framing of LLM evaluation, in both academic benchmarks and enterprise practice, treats it as a \textbf{model selection} exercise: run a suite of tests, rank the models, deploy the winner~\citep{liang2023helm, chang2024survey, zhang2024enterprise}. This framing captures only a fraction of the value evaluation can provide.

In production systems, the model is rarely the only---or even the primary---variable determining output quality. The prompt, retrieval pipeline, configuration parameters, and input data formatting all shape the result. When a system underperforms, the question is not ``which model should we switch to?'' but ``what's wrong and what should we change?'' Evaluation systems designed solely for model ranking cannot answer this question: they produce a score, not a diagnosis.

\subsection{Motivation}

Two developments motivate a rethinking of evaluation's role. First, continuous evaluation advocates observe that fixed benchmarks fall short for enterprise-scale agents where requirements evolve continuously~\citep{saxena2025continuous}, and that point-in-time analyses do not address companies' need to continuously assess tool reliability~\citep{azanza2025tracking}. These observations establish that evaluation must be ongoing---but ongoing measurement alone is insufficient if it does not produce actionable signals.

Second, the prompt engineering literature demonstrates that systematic, metric-driven iteration consistently outperforms one-shot design. \citet{sclar2024quantifying} show that minor prompt formatting changes can swing model performance by up to 76 percentage points. APE~\citep{zhou2023ape} and DSPy~\citep{khattab2023dspy} demonstrate that automated prompt refinement, guided by evaluation metrics, produces prompts that outperform human-designed alternatives. These findings suggest that evaluation's primary value lies not in selecting between models but in guiding the iterative improvement of the system around a model.

\subsection{Gap}

Despite these converging insights---that evaluation should be continuous~\citep{saxena2025continuous, azanza2025tracking} and that iterative refinement works~\citep{sclar2024quantifying, zhou2023ape, khattab2023dspy}---no existing methodology combines them into a coherent workflow. The continuous evaluation literature focuses on \textit{when} to re-evaluate but not on \textit{what to do} with the results beyond tracking quality over time. The prompt engineering literature optimizes prompts against aggregate metrics but does not address how to \textit{diagnose} which aspect of a prompt is causing failures. Multi-dimensional evaluation frameworks like HELM~\citep{liang2023helm} and DecodingTrust~\citep{wang2023decodingtrust} demonstrate that models have heterogeneous strength profiles across dimensions, but frame this as a reporting concern rather than a diagnostic tool for iterative improvement.

The result is a gap between evaluation-as-measurement and evaluation-as-improvement. Practitioners who want to use evaluation diagnostically must improvise: manually inspecting failure cases, guessing which system variable to change, and running ad hoc experiments without a structured workflow.

\subsection{Contributions}

We present \textbf{EvalLoop}, a methodology that reframes evaluation as a feedback loop for iterative system improvement. EvalLoop is organized around three core mechanisms:

\begin{enumerate}
\item \textbf{Dimensional metric grouping.} Metrics are grouped by business-relevant quality dimensions, enabling diagnosis of orthogonal failure modes. When Structural Compliance is 96\% but Hallucination Free Rate is 42\%, these are different root causes requiring different interventions---a distinction invisible in aggregate scoring.

\item \textbf{Failure mode classification.} For judge-evaluated dimensions, the evaluation system classifies \textit{why} outputs fail---not just \textit{that} they fail. This bridges the gap between dimensional diagnosis and actionable intervention.

\item \textbf{Iteration workflow.} A structured diagnose-hypothesize-intervene-measure cycle that treats evaluation runs as experiments, making the impact of each change visible and attributable.
\end{enumerate}

We validate EvalLoop through a case study on sales intelligence briefing generation (10 models, 3 providers, 18 metrics, 3 iterations). The methodology enabled a targeted prompt fix that improved the best model from 82.6\% to 94.6\% overall. We package EvalLoop as a reusable artifact bundle (practitioner playbook, coding agent specification, template repository).

\subsection{Paper Organization}

Section~\ref{sec:related} positions this work relative to LLM evaluation frameworks, judge reliability research, and prompt engineering. Section~\ref{sec:problem} defines the problem. Section~\ref{sec:methodology} presents the EvalLoop methodology. Section~\ref{sec:casestudy} validates through our case study. Section~\ref{sec:discussion} discusses generalizability and threats to validity. Section~\ref{sec:conclusion} concludes.

\section{Background and Related Work}
\label{sec:related}

\subsection{LLM Evaluation Frameworks}

The evaluation of large language models has evolved from single-metric benchmarks to multi-dimensional assessment frameworks. HELM~\citep{liang2023helm} established the principle of holistic evaluation across seven metric categories, demonstrating that models exhibit heterogeneous performance profiles. \citet{chang2024survey} provide a comprehensive taxonomy organizing evaluation along three axes: \textit{what} to evaluate, \textit{where}, and \textit{how}. DecodingTrust~\citep{wang2023decodingtrust} extends multi-dimensional evaluation to trustworthiness, assessing GPT models across eight dimensions and finding that high capability does not guarantee trustworthiness.

These frameworks share a common limitation: they evaluate models \textit{on} dimensions but do not prescribe how to \textit{use} dimensional results to improve the system. Our methodology addresses this gap by connecting dimensional evaluation to iteration.

For domain-specific NLG evaluation, G-Eval~\citep{liu2023geval} demonstrates that structured rubrics improve LLM judge correlation with human judgments. FActScore~\citep{min2023factscore} introduces atomic decomposition for factuality. We extend this principle into failure mode classification: not just identifying \textit{which} claims are unsupported, but categorizing \textit{why} they are unsupported to guide prompt fixes.

\subsection{LLM-as-Judge Reliability}

LLM judges have become the primary evaluation mechanism for semantic quality dimensions~\citep{zheng2023judging}. However, their reliability is contested. \citet{wang2023notfair} demonstrate position bias where response ordering affects rankings. \citet{panickssery2024self} show systematic self-preference: LLM evaluators favor their own family's generations. \citet{li2024prd} show that diverse judge panels outperform homogeneous ones. \citet{shankar2024validates} raise the meta-evaluation question of judge validation protocols.

Our methodology incorporates these findings through cross-provider judge panels (Section~\ref{sec:judges}): judges from at least two different model providers, with rubric-based prompts and multi-judge aggregation.

\subsection{Domain-Specific Enterprise Evaluation}

\citet{zhang2024enterprise} evaluate LLMs across enterprise-specific tasks and find that ``no model dominates across all tasks.'' The Sales Research Bench~\citep{bhol2025sales} evaluates sales AI across eight customer-weighted quality dimensions. Two recent papers argue for continuous evaluation: \citet{saxena2025continuous} propose continuous benchmark generation, and \citet{azanza2025tracking} present a framework for tracking evaluation as a ``moving target.''

Our work shares the continuous evaluation premise but extends it: evaluation should not only be ongoing but \textit{diagnostic}---producing signals that drive specific system changes.

\subsection{Prompt Engineering as Iterative Process}

The prompt engineering literature demonstrates that systematic optimization outperforms one-shot design. APE~\citep{zhou2023ape} shows that LLMs can generate prompts matching human engineer performance. DSPy~\citep{khattab2023dspy} compiles declarative LLM programs into optimized prompts via metric-driven iteration. \citet{sclar2024quantifying} quantify the stakes: minor formatting changes can swing performance by up to 76 percentage points.

These works optimize prompts against \textit{aggregate} metrics. None addresses the diagnostic question: which \textit{aspect} of the prompt is causing which \textit{type} of failure? EvalLoop fills this gap by connecting dimensional evaluation with failure mode classification to produce targeted prompt modifications.

\subsection{Positioning}

Table~\ref{tab:positioning} positions our contribution relative to the most closely related work.

\begin{table}[h]
\centering
\small
\begin{tabular}{lcccc}
\toprule
\textbf{Approach} & \textbf{Multi-dim.} & \textbf{Iterative} & \textbf{Diagnostic} & \textbf{Failure classif.} \\
\midrule
HELM~\citep{liang2023helm} & \checkmark & -- & -- & -- \\
FActScore~\citep{min2023factscore} & -- & -- & Partial & Claim-level \\
Sales Research Bench~\citep{bhol2025sales} & \checkmark & -- & -- & -- \\
Continuous Benchmarks~\citep{saxena2025continuous} & -- & \checkmark & -- & -- \\
DSPy / APE~\citep{khattab2023dspy, zhou2023ape} & -- & \checkmark & -- & -- \\
\textbf{EvalLoop (ours)} & \checkmark & \checkmark & \checkmark & \checkmark \\
\bottomrule
\end{tabular}
\caption{Positioning relative to related work.}
\label{tab:positioning}
\end{table}

\section{Problem: From Measurement to Diagnosis}
\label{sec:problem}

\subsection{Static Evaluation Misses the Primary Value}

Current LLM evaluation practice operates predominantly in what we term \textbf{model selection mode}: the system design is fixed, evaluation compares models, and the output is a ranking. Enterprise evaluation papers frame their contribution as helping organizations ``select the right model''~\citep{wang2025enterprise, zhang2024enterprise}. This framing misses the primary value of evaluation for deployed systems. Model selection is a one-time decision; system improvement is the continuous work.

\subsection{What Evaluation-Driven Improvement Requires}

For evaluation to serve as an improvement tool, three capabilities are necessary:

\textbf{Dimensional decomposition.} The evaluation must report \textit{where} the system is failing. An aggregate score of 82.6\% provides no diagnostic signal. A dimensional profile showing [Structural: 96\%, Content: 79\%, Hallucination: 85\%, Business Logic: 87\%, Synthesis: 66\%] immediately identifies improvement targets.

\textbf{Failure mode classification.} Within a weak dimension, the evaluation must report \textit{why} outputs fail. ``Hallucination rate is 85\%'' does not suggest a fix. ``69\% of hallucinations are inference-beyond-stated-facts'' directly implies a specific intervention.

\textbf{Iteration support.} The evaluation system must make re-evaluation cheap. Configuration-driven architecture, experiment tracking, and checkpoint recovery are infrastructure prerequisites.

\subsection{Why Dimensional Grouping Enables Diagnosis}

The key insight is that \textbf{different quality failures have different root causes and require different interventions}. Structural failures are caused by unclear format instructions. Hallucination failures are caused by missing grounding constraints. Synthesis failures are caused by absent paraphrasing requirements.

Aggregate scoring conflates these orthogonal failure modes. Dimensional grouping makes them distinguishable: if metrics are grouped by the intervention that would fix them, then identifying the weakest dimension is equivalent to identifying the most impactful next intervention.

This aligns with findings from multi-dimensional evaluation frameworks. \citet{zhang2024enterprise} demonstrate that models have heterogeneous profiles. HELM~\citep{liang2023helm} reports results across seven dimensions precisely because aggregate rankings obscure important distinctions. Our contribution connects this observation to a workflow: dimensional profiles are not just informative but \textit{actionable}.

\section{EvalLoop: A Design Methodology}
\label{sec:methodology}

EvalLoop is organized around six principles. Each is articulated independently of our case study and grounded in prior literature.

\subsection{Evaluation as Feedback Loop}

An evaluation system operates in one of two modes: (1)~\textbf{Model selection} (static)---the system design is fixed; evaluation compares models and picks the best one; and (2)~\textbf{System improvement} (iterative)---evaluation measures the impact of changes to system variables, enabling a diagnose-fix-measure cycle.

The evaluation literature overwhelmingly supports mode~1. Our methodology addresses mode~2. The iteration workflow proceeds as:

\begin{enumerate}
\item \textbf{Baseline evaluation.} Run all target models against the full metric suite. Obtain dimensional profiles.
\item \textbf{Diagnosis.} Identify the weakest dimension(s). Classify failure modes.
\item \textbf{Hypothesis.} ``Failures in dimension X are caused by system variable Y.''
\item \textbf{Intervention.} Change one system variable. Re-evaluate.
\item \textbf{Comparison.} Did the target dimension improve? Did others regress?
\end{enumerate}

The prompt is frequently the most productive variable to iterate on. \citet{sclar2024quantifying} demonstrate that minor prompt changes can swing performance by 10--70+ percentage points.

\begin{figure}[t]
\centering
\includegraphics[width=0.85\columnwidth]{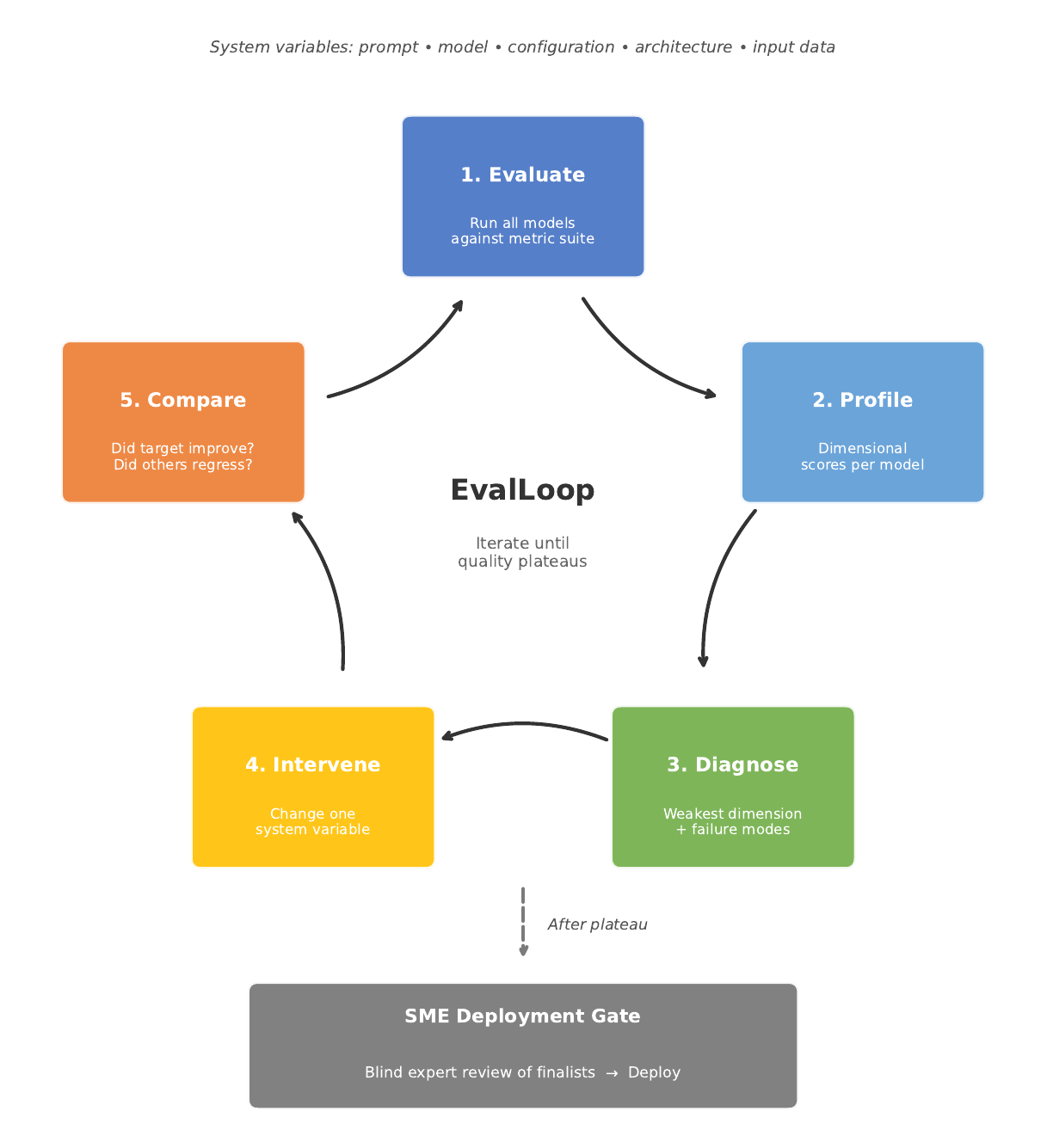}
\caption{The EvalLoop iteration cycle. System variables include prompt, model, configuration, architecture, and input data. After dimensional scores plateau, a one-time blind human gate (Section~\ref{sec:casestudy}, \S5.6) confirms deployment readiness.}
\label{fig:evalloop}
\end{figure}

\textbf{Final-stage human gate.} After the iteration loop plateaus and 3--5 finalists are short-listed by dimensional and operational criteria, we recommend a one-time blind review by a domain expert before deployment. The gate confirms that dimensional improvements correspond to perceived quality and resolves multi-criteria deployment trade-offs (cost, latency, provider diversity) that automated metrics cannot decide. Critically, the gate is a \textit{terminal step}, not a per-iteration check---automated metrics drive the hot loop, humans gate the cold one. This preserves iteration speed while preventing dimensional ceiling artifacts from masking residual quality issues. Section~\ref{sec:casestudy} (\S5.6) instantiates this on the case study. Figure~\ref{fig:evalloop} illustrates the full cycle and its terminal human gate.

\subsection{Dimensional Metric Grouping}

We recommend grouping metrics by \textbf{business-relevant quality dimension} rather than by failure severity or measurement technique. The rationale: dimensions aligned with distinct failure modes make diagnosis actionable.

\textbf{Definition.} A \textit{dimension} is a named set of metrics satisfying two criteria: (1)~the metrics test a common underlying quality aspect recognizable to stakeholders (\textit{communicational validity}), and (2)~the metrics plausibly share an intervention path (\textit{interventional validity}).

We recommend validating criterion~2 by checking \textit{intervention coherence}: when the system changes, do metrics within the dimension move in the same direction? In our case study, Content Accuracy showed 75\% intervention coherence; Business Logic showed 71\%.

Static within-dimension correlation (Pearson, phi coefficient) is \textit{not} a reliable validation criterion. Our empirical analysis found near-zero correlation within all dimensions---because many metrics are binary at ceiling ($>$88\% pass rates). Data-driven clustering produced groupings with no correspondence to expert dimensions (Adjusted Rand Index = $-$0.09). Dimensions serve \textit{diagnostic and communicational} purposes, not statistical ones.

\textbf{Recommended dimension count: 5--8.} Based on precedent from HELM (7), DecodingTrust (8), Sales Research Bench (8), and our case study (5).

\subsection{Prompt Iteration as Primary Workflow}

In our case study, the prompt---not the model---was the primary source of quality issues. The same model improved from 82.6\% to 94.6\% through a single targeted prompt change, while an undirected configuration change in a prior iteration produced zero quality impact.

\textbf{Failure mode classification} bridges diagnosis and action. Dimensional profiles tell a practitioner \textit{where} the system is weak; failure mode classification tells them \textit{why}. Categorizing 4,218 hallucination instances revealed: 41\% were inferences beyond stated facts, 28\% were claims neither confirmed nor denied, 20\% were misattributions, 2\% were direct contradictions. The dominant failure mode (69\%) pointed to a specific prompt weakness: encouraging synthesis without grounding constraints.

This extends FActScore's atomic decomposition~\citep{min2023factscore} and CritiqueLLM's critique generation~\citep{ke2024critiquellm} into a systematic classification step.

\subsection{Measurement Patterns: Deterministic Metrics and LLM Judges}
\label{sec:judges}

\begin{figure}[t]
\centering
\includegraphics[width=0.95\columnwidth]{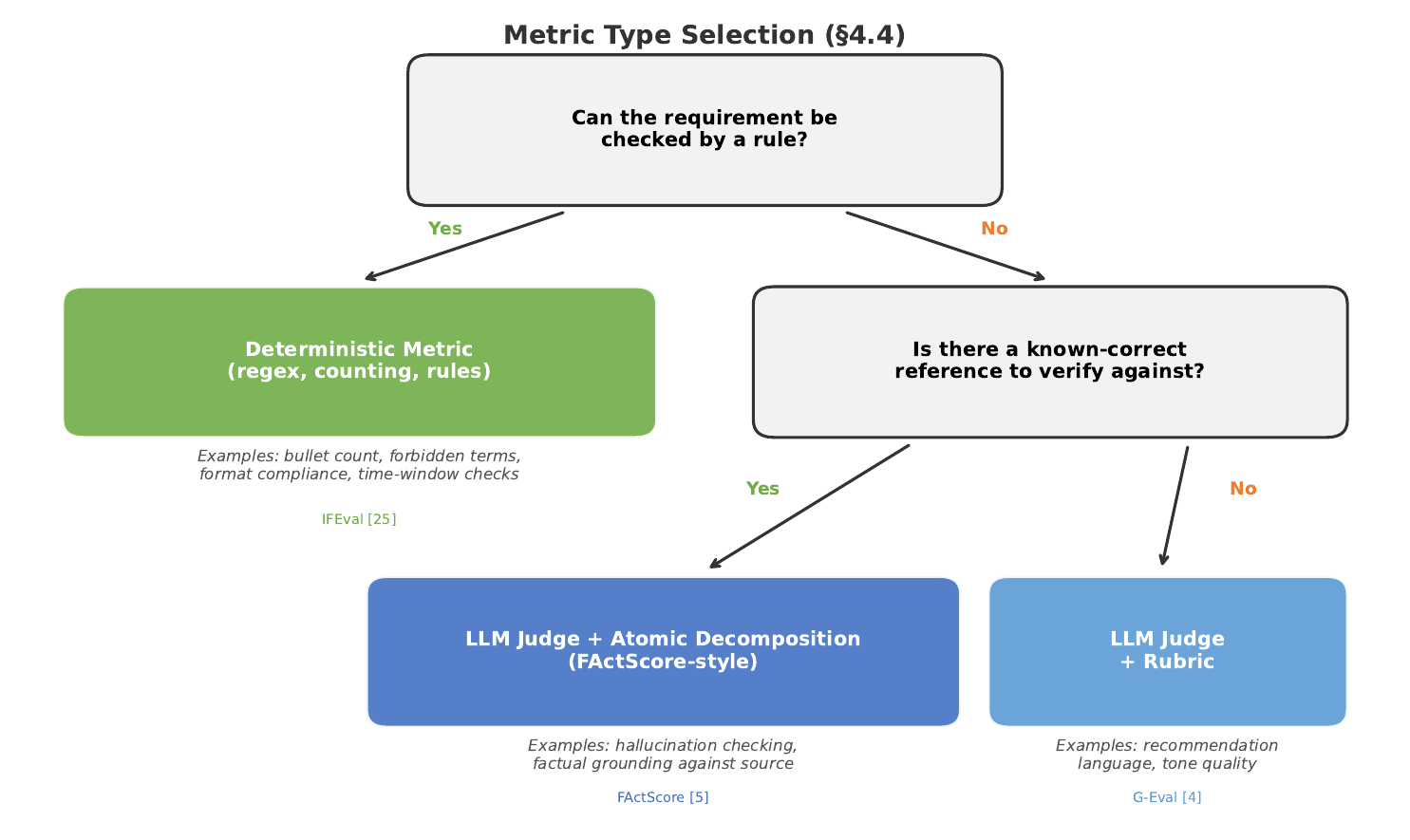}
\caption{Decision tree for metric type selection.}
\label{fig:decision_tree}
\end{figure}

We recommend a \textbf{deterministic-first heuristic}: when a requirement can be checked by a rule, prefer the rule over an LLM judge. Rules are cheap, reproducible, and immune to judge biases~\citep{wang2023notfair}. Figure~\ref{fig:decision_tree} illustrates the decision heuristic.

When LLM judges are necessary, their unreliability must be managed through: (1)~cross-provider panels~\citep{wang2023notfair, li2024prd}; (2)~rubric-based prompts~\citep{liu2023geval, kim2024prometheus}; (3)~multi-judge aggregation reporting mean, median, and consensus; (4)~failure mode classification in judge outputs; and (5)~meta-evaluation honesty---framing judges as heuristic, not ground truth~\citep{shankar2024validates}.

\subsection{Architectural Patterns for Iteration}

The evaluation system's architecture should make iteration cheap:

\begin{enumerate}
\item \textbf{Configuration-driven design.} Model lists, prompt versions, and thresholds live in YAML/JSON---not hardcoded. Following DSPy's separation principle~\citep{khattab2023dspy}.
\item \textbf{Protocol-based evaluators.} Each metric implements a common interface. New dimensions require new evaluator files, not orchestrator changes.
\item \textbf{Provider-agnostic inference.} An abstraction layer isolates provider-specific API logic.
\item \textbf{Batch processing with checkpoint recovery.} Long evaluations are resumable.
\item \textbf{Experiment tracking.} Every run logs to MLflow with full configuration and per-metric scores.
\end{enumerate}

\subsection{Developer vs.\ Agent Responsibilities}

\textbf{Decisions the developer must make:} Dimension definitions, metric thresholds, rubric content, when to stop iterating.

\textbf{Research the developer should conduct:} Available judge models, provider rate limits, existing benchmarks.

\textbf{Scaffolding a coding agent can generate:} Orchestrator skeleton, batch processor, experiment tracking integration, evaluator templates.

This three-way split enables a reusable artifact bundle: a \textit{playbook} guides developer decisions, an \textit{agent specification} tells a coding agent what to scaffold, and a \textit{template repository} provides the starting point.

\section{Case Study: Sales Intelligence Briefing Generation}
\label{sec:casestudy}

\subsection{Domain, Task, and Setup}

\textbf{Task.} Given a structured fact set describing a customer account, generate a concise bulleted briefing for a sales representative. The output must synthesize across facts, respect domain terminology rules, and avoid stating anything not grounded in the provided facts.

\textbf{Test corpus.} 100 synthetic account fact sets generated through a controlled pipeline: entity definitions are expanded into a matrix of candidate data points, populated with mock values, filtered via explicit rules, and emitted as normalized atomic key-value claims. The resulting fact sets exhibit realistic sparsity and domain variation while avoiding disclosure of proprietary data.

\textbf{Models evaluated.} 10 models across 3 providers: GPT-5.4, GPT-5.4-mini, GPT-5.4-nano (Azure OpenAI); Claude Opus 4.6, Claude Sonnet 4.6, Claude Haiku 4.5 (AWS Bedrock); Gemini 2.5 Pro, Gemini 2.5 Flash Lite, Gemini 3.1 Pro, Gemini 3.1 Flash Lite (Google Vertex AI).

\textbf{Configuration.} Temperature 0.0 (except Gemini 3.1 at provider default of 1.0). Max tokens: 2,000. All models received identical prompts via a shared prompt configuration. Judge panel: Claude Sonnet 4.6, GPT-5.4, Gemini 3.1 Pro.

\textbf{Metrics.} 18 metrics grouped into 5 dimensions: Structural Compliance (3 deterministic), Content Accuracy (4 deterministic + 1 judge), Hallucination Free Rate (1 judge, 3-judge panel), Business Logic (7 deterministic), Synthesis Power (2 deterministic).

\subsection{Applying the EvalLoop Methodology}

We executed three iterations:

\textbf{Iteration 1: Baseline (prompt v2.0).} The best model (gpt-5.4-nano, 87.4\% overall) scored unevenly across dimensions. gpt-5.4 ranked 4th at 82.6\%, with Synthesis Power at only 65.9\%.

\textbf{Iteration 2: Configuration experiment.} Disabled reasoning tokens for Gemini models. Result: no significant change in any dimension---illustrating undirected iteration.

\textbf{Iteration 3: Prompt refinement (v2.0 $\to$ v3.0).} Failure mode classification identified three actionable problems: (1)~forbidden-term ambiguity causing Content Accuracy failures; (2)~missing synthesis instruction causing high regurgitation; (3)~prompt encouraging inference without grounding constraints (69\% of hallucinations).

\subsection{Cross-Run Results}

Table~\ref{tab:gpt54-significance} shows the dimensional impact on gpt-5.4 with paired statistical tests.

\begin{table}[h]
\centering
\caption{gpt-5.4 dimensional profiles across prompt iterations with paired tests. CIs are 95\% bootstrap intervals (10{,}000 resamples, seed=42) on the per-case mean difference. p-values from two-sided paired $t$-tests; \textit{p (Bonf.)} applies Bonferroni correction across the five dimensions ($\alpha=0.01$). Wilcoxon signed-rank tests (not shown) agree with all $t$-test conclusions. Significance: *** $p<0.001$, ** $p<0.01$, * $p<0.05$, ns $=$ not significant.}
\label{tab:gpt54-significance}
\footnotesize
\begin{tabular}{lrrrrrrrrr}
\toprule
Dimension & n & v2.0 & v3.0 & $\Delta$ (pp) & 95\% CI (pp) & p (raw) & p (Bonf.) & Cohen's d & Sig. \\
\midrule
Structural Compliance & 100 & 96.3\% & 95.3\% & -1.0 & [-3.0, +1.0] & 0.368 & 1.000 & -0.09 & ns \\
Content Accuracy & 100 & 79.4\% & 96.2\% & +16.8 & [+14.2, +19.6] & $<\!0.001$ & $<\!0.001$ & +1.21 & *** \\
Hallucination Free Rate & 100 & 84.8\% & 93.6\% & +8.8 & [+4.4, +13.2] & $<\!0.001$ & $<\!0.001$ & +0.40 & *** \\
Business Logic & 100 & 86.5\% & 95.4\% & +8.9 & [+6.8, +11.1] & $<\!0.001$ & $<\!0.001$ & +0.83 & *** \\
Synthesis Power & 100 & 65.9\% & 92.3\% & +26.4 & [+23.0, +29.8] & $<\!0.001$ & $<\!0.001$ & +1.51 & *** \\
Overall (5-dim mean) & 100 & 82.6\% & 94.6\% & +12.0 & [+10.4, +13.5] & $<\!0.001$ & $<\!0.001$ & +1.53 & *** \\
\bottomrule
\end{tabular}
\end{table}

\textbf{Statistical methodology.} Because the same 100 test cases drive both runs, we use paired tests on per-case dimensional scores. We report two-sided paired $t$-tests (Bonferroni-corrected over the 5 dimensions, $\alpha_\text{adj}=0.01$), 95\% percentile bootstrap confidence intervals on the per-case mean difference (10{,}000 resamples, fixed seed for reproducibility), Wilcoxon signed-rank as a non-parametric robustness check, and Cohen's $d$ for paired samples as the effect size. Although Shapiro--Wilk rejects strict normality of the paired differences for several dimensions, the Central Limit Theorem applies at $n=100$, and bootstrap CIs match parametric t-CIs to within $0.4$pp on every dimension; we report both for transparency. Cross-dimension correlation of paired differences is small ($|r| \leq 0.21$), so Bonferroni's independence assumption is met and Holm--Bonferroni produces identical conclusions. Per-cell sample sizes vary slightly in the appendix (n = 97--100): rows where every judge call failed for a metric are dropped from that dimension's test only.

Improvement concentrated in dimensions targeted by the prompt change. Four of five dimensions show statistically significant gains after Bonferroni correction (Content Accuracy, Hallucination Free Rate, Business Logic, Synthesis Power; all $p<0.001$), with very large effect sizes for Content Accuracy ($d=1.21$) and Synthesis Power ($d=1.51$). The overall improvement of $+12.0$pp (95\% CI $[+10.4, +13.5]$) is highly significant with very large effect size ($d=1.53$).

\textbf{Structural Compliance: practical equivalence, not just non-significance.} Structural Compliance was not a target of the prompt revision. Its paired difference is statistically indistinguishable from zero ($p=0.37$, Wilcoxon $p=0.37$, Cohen's $d=-0.09$), and 89\% of per-case differences are exactly zero. The 95\% CI of $[-3.0, +1.0]$pp lies entirely within a $\pm 5$pp practical-equivalence band, so we read this as evidence of \emph{equivalence}, not inconclusiveness. This is the result the dimensional-profiling claim predicts: targeted prompt changes should improve diagnosed dimensions while leaving non-targeted dimensions practically unchanged.

\begin{figure}[t]
\centering
\includegraphics[width=0.9\columnwidth]{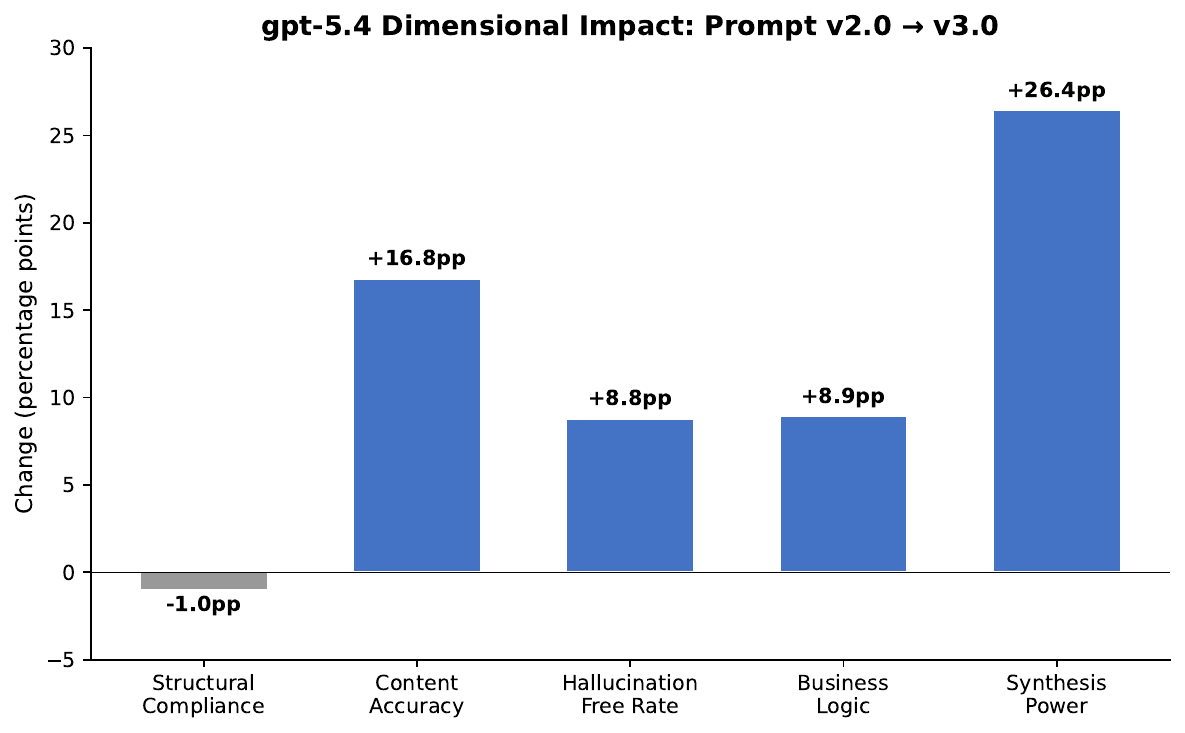}
\caption{Dimensional impact of prompt v2.0 $\to$ v3.0 on gpt-5.4.}
\label{fig:deltas}
\end{figure}

Figure~\ref{fig:deltas} visualizes the per-dimension impact. No single model dominates all dimensions (Table~\ref{tab:top5}): Gemini 2.5 Pro leads Synthesis (98.9\%) but trails Hallucination (85.0\%); gpt-5.4-mini leads Hallucination (95.0\%) with a 10pp advantage for hallucination-sensitive deployments.

\begin{table}[h]
\centering
\small
\begin{tabular}{lccccc|c}
\toprule
\textbf{Model} & \textbf{Str.} & \textbf{Cont.} & \textbf{Hal.} & \textbf{Biz} & \textbf{Syn.} & \textbf{Overall} \\
\midrule
GPT-5.4 & 95.3 & 96.2 & 93.6 & 95.4 & 92.3 & \textbf{94.6} \\
Gemini 2.5 Pro & 99.3 & 94.0 & 85.0 & 89.7 & 98.9 & 93.4 \\
Gemini 2.5 Flash & 94.3 & 97.0 & 91.2 & 90.4 & 89.7 & 92.5 \\
GPT-5.4-mini & 95.0 & 93.4 & \textbf{95.0} & 91.3 & 87.8 & 92.5 \\
Claude Opus 4.6 & 98.3 & 92.2 & 90.4 & 82.5 & 94.6 & 91.6 \\
\bottomrule
\end{tabular}
\caption{Top 5 models by 5-dimension mean (v3.0). Bold highlights best-in-column.}
\label{tab:top5}
\end{table}

\subsection{What Dimensional Analysis Revealed}

\textbf{Provider-level patterns.} Claude models averaged 84.3\% Hallucination but 95.8\% Synthesis; GPT models averaged 91.3\% Hallucination but 89.8\% Synthesis---suggesting systematic differences in prompt interpretation invisible to aggregate scoring.

\textbf{Framing comparison.} For hallucination-sensitive deployment, the aggregate-ranked winner (Gemini 2.5 Pro, 93.4\%) has only 85.0\% hallucination score. The dimensionally-informed choice (GPT-5.4-mini, 95.0\% hallucination) provides a 10pp safety advantage at lower cost.

\textbf{Inter-judge agreement.} Mean pairwise $r$=0.51 on hallucination, 67.4\% unanimous agreement. Recommendation language achieved $r$=0.65, 99.4\% unanimous---consistent with the deterministic-first heuristic.

\subsection{Dimensional Validation}

We validated the 5-dimension grouping using phi coefficients, hierarchical clustering, cross-run delta coherence, and mutual information. Content Accuracy showed moderate validation (75\% intervention coherence, 3.3$\times$ MI ratio). Business Logic showed no statistical cohesion (ARI=$-$0.09 vs.\ data-driven clustering). This finding is itself diagnostic: Business Logic is communicationally valid but not interventionally valid---its metrics require per-rule interventions rather than a single fix. Three metrics saturate at 100\% pass in this run (\texttt{bullet\_format\_compliance}, \texttt{contact\_age\_filter}, \texttt{length\_final}) and are excluded from the binary-association tracks because phi and mutual information are undefined for zero-variance variables; they remain part of their respective dimensions and re-enter the intervention-coherence analysis.

\begin{figure}[t]
\centering
\includegraphics[width=0.85\columnwidth]{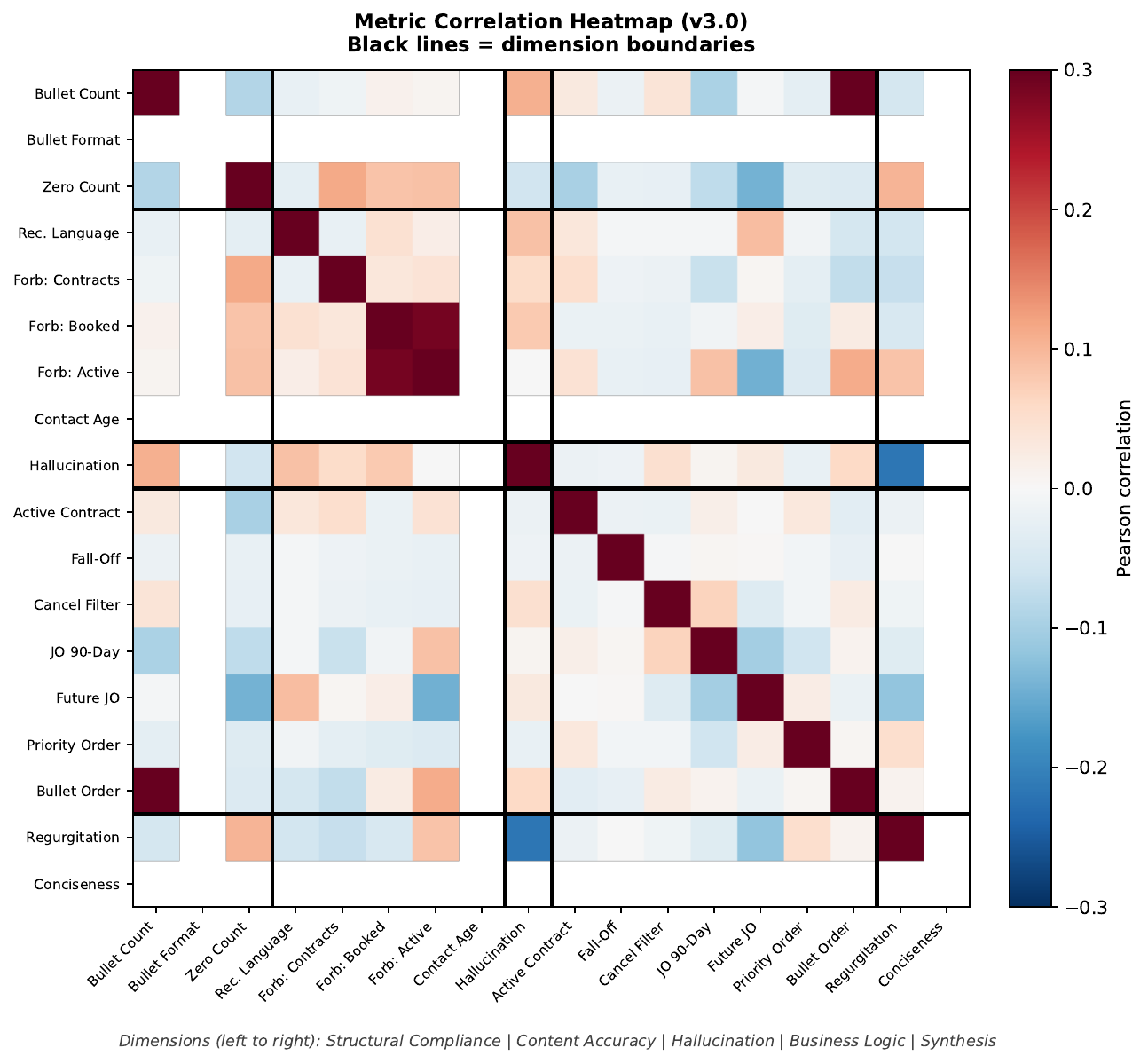}
\caption{Metric correlation heatmap (v3.0). Near-zero within-dimension correlation confirms dimensions are conceptual groupings, not statistical clusters.}
\label{fig:heatmap}
\end{figure}

\subsection{Human Validation: SME Deployment Gate}
\label{sec:human_gate}

After Iteration~3, four finalists were short-listed from the 10 evaluated models by combining dimensional scores, cost, latency, and provider diversity: gpt-5.4, gpt-5.4-mini, gemini-2.5-flash-lite, and Claude Opus 4.6. A subject-matter expert with experience in the target sales role then performed a blind review on a 16-case sample chosen to span all business segments and contract types. Per case, the SME received four outputs labeled ``Model 1''--``Model 4'' with the model-to-label mapping randomized, and selected the \textit{best} and \textit{2nd-best}. Per-model preference was summarized via a weighted count ($0.7 \times \text{best} + 0.3 \times \text{2nd-best}$).

\begin{table}[h]
\centering
\small
\begin{tabular}{lccc}
\toprule
\textbf{Model} & \textbf{Best} & \textbf{2nd-Best} & \textbf{Weighted} \\
\midrule
Claude Opus 4.6 & 5 & 8 & \textbf{5.9} \\
gpt-5.4-mini & 6 & 3 & 5.1 \\
gpt-5.4 & 4 & 1 & 3.1 \\
gemini-2.5-flash-lite & 1 & 2 & 1.3 \\
\bottomrule
\end{tabular}
\caption{SME blind preference (16 cases $\times$ 4 finalists). Weighted = $0.7 \times \text{best} + 0.3 \times \text{2nd-best}$.}
\label{tab:sme}
\end{table}

Claude Opus 4.6 was the SME's top weighted choice (top-2 in 13 of 16 cases, 81\%) and was selected for deployment after combining the SME signal with cost and latency. The dimensional ranking and the human gate converged: gemini-2.5-flash-lite, weakest on dimensional scores, was also weakest in human review, while opus and gpt-5.4-mini---both strong dimensionally---were the two human-preferred finalists.

\textbf{Convergence with EvalLoop dimensions.} SME free-text notes (12 of 16 cases) clustered cleanly into the existing dimensional taxonomy: length/concision concerns (Synthesis Power), role-inappropriate terminology such as ``Active'' used in Perm contracts (Content Accuracy), priority-ordering observations (Business Logic), and spurious content for absent facts (Hallucination Free Rate). No novel failure mode emerged in human review---the dimensional taxonomy held up under blind expert scrutiny.

\textbf{Iteration economy.} Human review focused on a 4-model $\times$ 16-case sample (64 observations) instead of the full 10-model $\times$ 100-case design (1,000 observations)---a 94\% reduction in review burden---without losing validation strength: the dimensional profiles had already filtered out clearly-failing candidates (no finalist received a ``None Are Great'' verdict on any case).

\textbf{Limitations.} Single-reviewer design with no inter-rater agreement; 16 of 100 cases reviewed; rankings rather than per-dimension numeric scores. The gate is positioned as a deployment checkpoint, not a calibration of LLM judges against human ground truth---that is a separate study.

\section{Discussion}
\label{sec:discussion}

\subsection{Generalizability}

EvalLoop's principles apply to any LLM-powered system where output quality decomposes into distinguishable dimensions, different failure modes suggest different interventions, and the practitioner can re-evaluate after changes. The specific dimensions are task-dependent; the methodology prescribes the \textit{process} of deriving, validating, and iterating on dimensions.

The case study's finding that the prompt was the primary bottleneck may not generalize to all domains. For reasoning-heavy tasks, model capability may be the limiting factor. For retrieval-augmented systems, retrieval quality may dominate. EvalLoop supports iterating on any system variable; the prompt-first heuristic is a recommended starting point, not an invariant.

\subsection{Boundary Conditions}

EvalLoop is not appropriate when: output has no decomposable structure (open-ended creative writing); the system is already at ceiling on all dimensions; the bottleneck is outside the evaluation loop (training data, retrieval quality); or evaluation cost is prohibitive for rapid iteration.

\subsection{Threats to Validity}

\textbf{Internal validity.} The iteration improvement was measured on the same 100-account test corpus. Overfitting to the test distribution remains a risk absent a held-out validation set. Statistical testing on per-case paired deltas (Table~\ref{tab:gpt54-significance}) confirms that four of five dimensional improvements survive Bonferroni correction with medium-to-very-large effect sizes; the Structural Compliance change ($-$1.0pp) is reported transparently as non-significant.

\textbf{External validity.} Results derive from a single task and domain. Generalization is unvalidated, though methodology principles are grounded in multi-domain literature.

\textbf{Construct validity.} The 5-dimension grouping was expert-defined, not data-derived. Two dimensions showed genuine statistical structure; others group independent metrics. We report this transparently.

\textbf{Judge reliability.} Inter-judge agreement ($r$=0.51, 67.4\% unanimous) indicates moderate reliability. Judges remain heuristic evaluators whose agreement with human ground truth was not validated quantitatively in this study. The SME gate (\S\ref{sec:human_gate}) provides qualitative confirmation---human-preferred finalists matched dimensional rankings and SME concerns mapped onto the existing dimensions---but a single reviewer on 16 cases is not a calibration of judges against humans.

\textbf{Human gate scope.} The SME gate complements LLM-judge evaluation rather than replacing it: judges scale to 1,000-evaluation iterations; humans gate before deployment. We position human review as a one-time terminal step to preserve iteration speed. Higher-stakes domains (regulated communications, medical, legal) may need a larger panel and per-iteration sampling---the methodology supports this, but our case study does not validate it.

\section{Conclusion}
\label{sec:conclusion}

We presented EvalLoop, a methodology for evaluation-driven iterative improvement of business AI systems. EvalLoop reframes evaluation from static model selection to a diagnostic feedback loop: dimensional metric grouping surfaces \textit{where} the system is failing, failure mode classification reveals \textit{why}, and a structured iteration workflow translates diagnosis into targeted fixes with measurable impact.

Our case study demonstrated the methodology's value: dimensional diagnosis identified prompt-induced interpretation errors as the dominant hallucination source, enabling a targeted fix that improved the best model from 82.6\% to 94.6\%. The empirical validation of dimensional grouping produced a finding of independent interest: expert-defined dimensions serve communication and intervention-targeting purposes but do not necessarily correspond to statistical structure in the data.

\textbf{Future work.} Longitudinal validation across multiple domains; automated dimension discovery from requirements; integration with automated prompt optimization where dimensional diagnosis guides the optimizer's objective function.

\section*{Acknowledgements}

This work was conducted at Robert Half, where the Data Science team builds, deploys, and evaluates production AI systems for enterprise use. We thank Robert Half for encouraging the publication of applied research methodologies developed in the course of this work and for fostering an environment that supports experimentation, evaluation, and continuous improvement of AI systems.

We are grateful to the colleagues, practitioners, and business stakeholders whose domain expertise and feedback helped shape the evaluation framework and case study presented in this paper. Their collaboration helped ensure that the methodology addressed practical challenges encountered in real-world enterprise AI deployments.

\bibliographystyle{plainnat}
\bibliography{references}

\appendix
\section{All-Models Significance Results}
\label{sec:appendix-significance}

Table~\ref{tab:all-models-significance} extends the gpt-5.4 analysis (Table~\ref{tab:gpt54-significance}) to all 10 evaluated models, using the same paired-test methodology (paired $t$-tests with Bonferroni correction over 5 dimensions, 95\% bootstrap CIs with 10{,}000 resamples and fixed seed, Wilcoxon as a robustness check, and Cohen's $d$). Per-cell sample sizes vary by 0--3 because rows where every judge call failed for a metric are dropped from that dimension's test only: the GPT and Gemini~2.5 families retain $n=100$ on every dimension; gemini-3.1-pro drops to $n=99$ on Hallucination Free Rate; Claude Haiku and Claude Sonnet retain $n=99$ across all dimensions; Claude Opus retains $n=97$. The pattern is consistent across providers: Content Accuracy and Synthesis Power show large, highly significant gains for nearly every model; Hallucination Free Rate gains are largest for the Claude family (whose v2.0 baselines were lowest); Structural Compliance changes are mixed in sign and small in magnitude.

\begin{landscape}
\scriptsize
\begin{longtable}{llrrrrrrrrrr}
\caption{Significance-tested v2.0 → v3.0 dimensional improvements for all 10 evaluated models. Same statistical methodology as Table~\ref{tab:gpt54-significance}. Bedrock model IDs shortened (e.g., \texttt{us.anthropic.} prefix and version suffixes removed).}\label{tab:all-models-significance} \\
\toprule
Model & Dimension & n & v2.0 & v3.0 & $\Delta$ (pp) & 95\% CI (pp) & p (raw) & p (Bonf.) & Wilcoxon p & Cohen's d & Sig. \\
\midrule
\endfirsthead

\multicolumn{12}{c}{\tablename\ \thetable\ -- continued from previous page} \\
\toprule
Model & Dimension & n & v2.0 & v3.0 & $\Delta$ (pp) & 95\% CI (pp) & p (raw) & p (Bonf.) & Wilcoxon p & Cohen's d & Sig. \\
\midrule
\endhead

\midrule
\multicolumn{12}{r}{Continued on next page} \\
\endfoot

\bottomrule
\endlastfoot

gemini-2.5-flash-lite & Structural Compliance & 100 & 83.3\% & 94.3\% & +11.0 & [+7.3, +14.3] & $<\!0.001$ & $<\!0.001$ & $<\!0.001$ & +0.60 & *** \\
gemini-2.5-flash-lite & Content Accuracy & 100 & 62.2\% & 97.0\% & +34.8 & [+30.4, +39.4] & $<\!0.001$ & $<\!0.001$ & $<\!0.001$ & +1.52 & *** \\
gemini-2.5-flash-lite & Hallucination Free Rate & 100 & 65.5\% & 91.2\% & +25.7 & [+16.9, +34.7] & $<\!0.001$ & $<\!0.001$ & $<\!0.001$ & +0.56 & *** \\
gemini-2.5-flash-lite & Business Logic & 100 & 83.1\% & 90.4\% & +7.3 & [+4.5, +10.1] & $<\!0.001$ & $<\!0.001$ & $<\!0.001$ & +0.50 & *** \\
gemini-2.5-flash-lite & Synthesis Power & 100 & 90.4\% & 89.7\% & -0.8 & [-3.1, +1.5] & 0.514 & 1.000 & 0.643 & -0.07 & ns \\
gemini-2.5-flash-lite & Overall (5-dim mean) & 100 & 76.9\% & 92.5\% & +15.6 & [+12.6, +18.6] & $<\!0.001$ & $<\!0.001$ & $<\!0.001$ & +1.02 & *** \\
gemini-2.5-pro & Structural Compliance & 100 & 75.3\% & 99.3\% & +24.0 & [+21.0, +26.7] & $<\!0.001$ & $<\!0.001$ & $<\!0.001$ & +1.60 & *** \\
gemini-2.5-pro & Content Accuracy & 100 & 82.4\% & 94.0\% & +11.6 & [+8.9, +14.3] & $<\!0.001$ & $<\!0.001$ & $<\!0.001$ & +0.85 & *** \\
gemini-2.5-pro & Hallucination Free Rate & 100 & 67.6\% & 85.0\% & +17.4 & [+10.2, +24.6] & $<\!0.001$ & $<\!0.001$ & $<\!0.001$ & +0.48 & *** \\
gemini-2.5-pro & Business Logic & 100 & 75.8\% & 89.7\% & +13.9 & [+11.4, +16.4] & $<\!0.001$ & $<\!0.001$ & $<\!0.001$ & +1.11 & *** \\
gemini-2.5-pro & Synthesis Power & 100 & 99.1\% & 98.9\% & -0.2 & [-0.7, +0.3] & 0.469 & 1.000 & 0.062 & -0.07 & ns \\
gemini-2.5-pro & Overall (5-dim mean) & 100 & 80.1\% & 93.4\% & +13.3 & [+11.5, +15.2] & $<\!0.001$ & $<\!0.001$ & $<\!0.001$ & +1.44 & *** \\
gemini-3.1-flash-lite & Structural Compliance & 100 & 95.3\% & 98.0\% & +2.7 & [+0.3, +5.3] & 0.032 & 0.159 & 0.033 & +0.22 & ns \\
gemini-3.1-flash-lite & Content Accuracy & 100 & 79.0\% & 91.3\% & +12.3 & [+9.4, +15.0] & $<\!0.001$ & $<\!0.001$ & $<\!0.001$ & +0.87 & *** \\
gemini-3.1-flash-lite & Hallucination Free Rate & 100 & 73.9\% & 72.6\% & -1.3 & [-6.7, +4.1] & 0.647 & 1.000 & 0.764 & -0.05 & ns \\
gemini-3.1-flash-lite & Business Logic & 100 & 84.9\% & 90.9\% & +6.0 & [+3.7, +8.3] & $<\!0.001$ & $<\!0.001$ & $<\!0.001$ & +0.51 & *** \\
gemini-3.1-flash-lite & Synthesis Power & 100 & 98.5\% & 98.6\% & +0.0 & [-0.5, +0.5] & 0.940 & 1.000 & 0.897 & +0.01 & ns \\
gemini-3.1-flash-lite & Overall (5-dim mean) & 100 & 86.3\% & 90.3\% & +3.9 & [+2.6, +5.3] & $<\!0.001$ & $<\!0.001$ & $<\!0.001$ & +0.57 & *** \\
gemini-3.1-pro & Structural Compliance & 100 & 66.7\% & 78.0\% & +11.3 & [+8.0, +14.7] & $<\!0.001$ & $<\!0.001$ & $<\!0.001$ & +0.66 & *** \\
gemini-3.1-pro & Content Accuracy & 100 & 83.0\% & 93.2\% & +10.2 & [+7.2, +13.0] & $<\!0.001$ & $<\!0.001$ & $<\!0.001$ & +0.69 & *** \\
gemini-3.1-pro & Hallucination Free Rate & 99 & 67.7\% & 82.3\% & +14.6 & [+5.4, +24.1] & 0.002 & 0.012 & 0.001 & +0.31 & * \\
gemini-3.1-pro & Business Logic & 100 & 73.2\% & 80.4\% & +7.2 & [+4.9, +9.5] & $<\!0.001$ & $<\!0.001$ & $<\!0.001$ & +0.61 & *** \\
gemini-3.1-pro & Synthesis Power & 100 & 98.3\% & 97.4\% & -0.8 & [-2.5, +0.7] & 0.300 & 1.000 & 0.229 & -0.10 & ns \\
gemini-3.1-pro & Overall (5-dim mean) & 100 & 77.8\% & 86.3\% & +8.5 & [+6.3, +10.7] & $<\!0.001$ & $<\!0.001$ & $<\!0.001$ & +0.76 & *** \\
gpt-5.4 & Structural Compliance & 100 & 96.3\% & 95.3\% & -1.0 & [-3.0, +1.0] & 0.368 & 1.000 & 0.366 & -0.09 & ns \\
gpt-5.4 & Content Accuracy & 100 & 79.4\% & 96.2\% & +16.8 & [+14.2, +19.6] & $<\!0.001$ & $<\!0.001$ & $<\!0.001$ & +1.21 & *** \\
gpt-5.4 & Hallucination Free Rate & 100 & 84.8\% & 93.6\% & +8.8 & [+4.4, +13.2] & $<\!0.001$ & $<\!0.001$ & $<\!0.001$ & +0.40 & *** \\
gpt-5.4 & Business Logic & 100 & 86.5\% & 95.4\% & +8.9 & [+6.8, +11.1] & $<\!0.001$ & $<\!0.001$ & $<\!0.001$ & +0.83 & *** \\
gpt-5.4 & Synthesis Power & 100 & 65.9\% & 92.3\% & +26.4 & [+23.0, +29.8] & $<\!0.001$ & $<\!0.001$ & $<\!0.001$ & +1.51 & *** \\
gpt-5.4 & Overall (5-dim mean) & 100 & 82.6\% & 94.6\% & +12.0 & [+10.4, +13.5] & $<\!0.001$ & $<\!0.001$ & $<\!0.001$ & +1.53 & *** \\
gpt-5.4-mini & Structural Compliance & 100 & 96.0\% & 95.0\% & -1.0 & [-3.0, +1.0] & 0.368 & 1.000 & 0.366 & -0.09 & ns \\
gpt-5.4-mini & Content Accuracy & 100 & 77.2\% & 93.4\% & +16.2 & [+13.2, +19.2] & $<\!0.001$ & $<\!0.001$ & $<\!0.001$ & +1.08 & *** \\
gpt-5.4-mini & Hallucination Free Rate & 100 & 90.2\% & 95.0\% & +4.8 & [+2.0, +7.6] & 0.001 & 0.007 & 0.003 & +0.33 & ** \\
gpt-5.4-mini & Business Logic & 100 & 84.8\% & 91.3\% & +6.4 & [+4.3, +8.6] & $<\!0.001$ & $<\!0.001$ & $<\!0.001$ & +0.59 & *** \\
gpt-5.4-mini & Synthesis Power & 100 & 86.5\% & 87.8\% & +1.3 & [-0.6, +3.3] & 0.190 & 0.950 & 0.396 & +0.13 & ns \\
gpt-5.4-mini & Overall (5-dim mean) & 100 & 86.9\% & 92.5\% & +5.6 & [+4.4, +6.7] & $<\!0.001$ & $<\!0.001$ & $<\!0.001$ & +0.95 & *** \\
gpt-5.4-nano & Structural Compliance & 100 & 96.3\% & 92.3\% & -4.0 & [-6.3, -2.0] & $<\!0.001$ & 0.002 & $<\!0.001$ & -0.37 & ** \\
gpt-5.4-nano & Content Accuracy & 100 & 83.0\% & 97.0\% & +14.0 & [+11.5, +16.4] & $<\!0.001$ & $<\!0.001$ & $<\!0.001$ & +1.11 & *** \\
gpt-5.4-nano & Hallucination Free Rate & 100 & 83.4\% & 85.4\% & +2.0 & [-2.6, +6.4] & 0.390 & 1.000 & 0.284 & +0.09 & ns \\
gpt-5.4-nano & Business Logic & 100 & 87.0\% & 91.4\% & +4.4 & [+2.1, +6.6] & $<\!0.001$ & 0.002 & $<\!0.001$ & +0.37 & ** \\
gpt-5.4-nano & Synthesis Power & 100 & 87.3\% & 89.2\% & +1.9 & [-0.6, +4.5] & 0.154 & 0.769 & 0.775 & +0.14 & ns \\
gpt-5.4-nano & Overall (5-dim mean) & 100 & 87.4\% & 91.1\% & +3.6 & [+2.2, +5.1] & $<\!0.001$ & $<\!0.001$ & $<\!0.001$ & +0.49 & *** \\
claude-haiku-4-5 & Structural Compliance & 99 & 96.6\% & 98.3\% & +1.7 & [+0.3, +3.4] & 0.025 & 0.123 & 0.034 & +0.23 & ns \\
claude-haiku-4-5 & Content Accuracy & 99 & 73.2\% & 96.6\% & +23.4 & [+20.3, +26.5] & $<\!0.001$ & $<\!0.001$ & $<\!0.001$ & +1.50 & *** \\
claude-haiku-4-5 & Hallucination Free Rate & 99 & 23.4\% & 79.0\% & +55.6 & [+49.1, +61.8] & $<\!0.001$ & $<\!0.001$ & $<\!0.001$ & +1.73 & *** \\
claude-haiku-4-5 & Business Logic & 99 & 81.5\% & 83.3\% & +1.7 & [-0.5, +4.0] & 0.138 & 0.692 & 0.136 & +0.15 & ns \\
claude-haiku-4-5 & Synthesis Power & 99 & 93.0\% & 96.6\% & +3.6 & [+2.1, +5.2] & $<\!0.001$ & $<\!0.001$ & $<\!0.001$ & +0.46 & *** \\
claude-haiku-4-5 & Overall (5-dim mean) & 99 & 73.5\% & 90.8\% & +17.2 & [+15.6, +18.8] & $<\!0.001$ & $<\!0.001$ & $<\!0.001$ & +2.10 & *** \\
claude-opus-4-6-v1 & Structural Compliance & 97 & 95.2\% & 98.3\% & +3.1 & [+1.4, +5.2] & 0.002 & 0.011 & 0.003 & +0.32 & * \\
claude-opus-4-6-v1 & Content Accuracy & 97 & 69.7\% & 92.4\% & +22.7 & [+19.2, +26.3] & $<\!0.001$ & $<\!0.001$ & $<\!0.001$ & +1.24 & *** \\
claude-opus-4-6-v1 & Hallucination Free Rate & 97 & 41.9\% & 90.7\% & +48.9 & [+43.7, +54.0] & $<\!0.001$ & $<\!0.001$ & $<\!0.001$ & +1.89 & *** \\
claude-opus-4-6-v1 & Business Logic & 97 & 81.9\% & 82.5\% & +0.7 & [-1.5, +2.8] & 0.541 & 1.000 & 0.694 & +0.06 & ns \\
claude-opus-4-6-v1 & Synthesis Power & 97 & 66.6\% & 94.5\% & +27.9 & [+24.0, +31.6] & $<\!0.001$ & $<\!0.001$ & $<\!0.001$ & +1.48 & *** \\
claude-opus-4-6-v1 & Overall (5-dim mean) & 97 & 71.0\% & 91.7\% & +20.6 & [+18.9, +22.4] & $<\!0.001$ & $<\!0.001$ & $<\!0.001$ & +2.36 & *** \\
claude-sonnet-4-6 & Structural Compliance & 99 & 91.6\% & 96.3\% & +4.7 & [+2.7, +7.1] & $<\!0.001$ & $<\!0.001$ & $<\!0.001$ & +0.40 & *** \\
claude-sonnet-4-6 & Content Accuracy & 99 & 65.8\% & 94.1\% & +28.4 & [+24.9, +32.0] & $<\!0.001$ & $<\!0.001$ & $<\!0.001$ & +1.55 & *** \\
claude-sonnet-4-6 & Hallucination Free Rate & 99 & 15.5\% & 83.2\% & +67.7 & [+62.9, +72.4] & $<\!0.001$ & $<\!0.001$ & $<\!0.001$ & +2.79 & *** \\
claude-sonnet-4-6 & Business Logic & 99 & 78.7\% & 83.6\% & +4.9 & [+2.5, +7.3] & $<\!0.001$ & $<\!0.001$ & $<\!0.001$ & +0.40 & *** \\
claude-sonnet-4-6 & Synthesis Power & 99 & 56.0\% & 96.1\% & +40.2 & [+37.5, +42.5] & $<\!0.001$ & $<\!0.001$ & $<\!0.001$ & +3.15 & *** \\
claude-sonnet-4-6 & Overall (5-dim mean) & 99 & 61.5\% & 90.7\% & +29.2 & [+27.6, +30.7] & $<\!0.001$ & $<\!0.001$ & $<\!0.001$ & +3.59 & *** \\
\end{longtable}
\end{landscape}

\end{document}